\documentclass{article}

\usepackage{arxiv}

\usepackage[utf8]{inputenc} 
\usepackage[T1]{fontenc}    
\usepackage{hyperref}       
\usepackage{url}            
\usepackage{booktabs}       
\usepackage{amsfonts}       
\usepackage{nicefrac}       
\usepackage{microtype}      
\usepackage[title]{appendix}
\usepackage{graphicx}
\usepackage[bf]{caption}

\usepackage{pdflscape}      
\usepackage{booktabs}
\usepackage{longtable}
\usepackage{array}

\usepackage{doi}
\usepackage{etoolbox}
\usepackage[natbibapa]{apacite}
\AtBeginDocument{}
\renewcommand{\APACrefnote}[1]{}

\newtoggle{bibdoi}
\newtoggle{biburl}
\makeatletter
\newsavebox{\bib@url}
\newsavebox{\bib@doi}

\undef{\APACrefURL}
\undef{\endAPACrefURL}
\undef{\APACrefDOI}
\undef{\endAPACrefDOI}

\newenvironment{APACrefURL}
  {\global\toggletrue{biburl}\lrbox\bib@url}
  {\endlrbox}

\newenvironment{APACrefDOI}
  {\global\toggletrue{bibdoi}\lrbox\bib@doi}
  {\endlrbox}

\newcommand{\printinfo}{
  \iftoggle{bibdoi}{\usebox{\bib@doi}}{\usebox{\bib@url}}
  \togglefalse{bibdoi}
}

\AtBeginEnvironment{thebibliography}{
\pretocmd{\PrintBackRefs}{%
  \iftoggle{bibdoi}
    {\iftoggle{biburl}{\unskip\unskip}{}\usebox{\bib@doi}}
    {\iftoggle{biburl}{Retrieved from \usebox{\bib@url}}}{}
  \togglefalse{bibdoi}\togglefalse{biburl}%
}{}{}}
\makeatother

\title{Roadmap Towards Responsible AI in Crisis Resilience Management}

\date{} 					

\renewcommand{\headeright}{}
\renewcommand{\undertitle}{}
\renewcommand{\shorttitle}{}

\hypersetup{
pdftitle={Roadmap Towards Responsible AI in Crisis Resilience Management},
pdfsubject={},
}

\begin{document}
\maketitle

\begin{center}
{\Large
Cheng-Chun Lee\textsuperscript{a,*},
Tina Comes\textsuperscript{b}, 
Megan Finn\textsuperscript{c}, 
Ali Mostafavi\textsuperscript{a}
\par}

\bigskip
\textsuperscript{a} Urban Resilience.AI Lab, Zachry Department of Civil and Environmental Engineering,\\ Texas A\&M University, College Station, TX\\
\vspace{6pt}
\textsuperscript{b} Faculty of Technology, Policy and Management, Delft University of Technology,\\ Jaffalaan 5, 2628 BX, Delft, the Netherlands\\
\vspace{6pt}
\textsuperscript{c} Information School, University of Washington, Seattle, WA\\
\vspace{6pt}
\textsuperscript{*} correseponding author, email: ccbarrylee@tamu.edu
\\
\end{center}
\bigskip
\begin{abstract}
Novel data sensing and AI technologies are finding practical use in the analysis of crisis resilience, revealing the need to consider how responsible artificial intelligence (AI) practices can mitigate harmful outcomes and protect vulnerable populations. In this paper, we present a responsible AI roadmap that is embedded in the Crisis Information Management Circle. This roadmap includes six propositions to highlight and address important challenges and considerations specifically related to responsible AI for crisis resilience management. We cover a wide spectrum of interwoven challenges and considerations pertaining to the responsible collection, analysis, sharing, and use of information such as equity, fairness, biases, explainability and transparency, accountability, privacy and security, inter-organizational coordination, and public engagement. Through examining issues around AI systems for crisis resilience management, we dissect the inherent complexities of information management and decision-making in crises and highlight the urgency of responsible AI research and practice. The ideas laid out in this paper are the first attempt in establishing a roadmap for researchers, practitioners, developers, emergency managers, humanitarian organizations, and public officials to address important considerations for responsible AI pertaining to crisis resilience management.
\end{abstract}



\section{Introduction}
Crises resulting from natural hazards (e.g., flooding, hurricanes, wildfires, earthquakes), pandemics, and conflicts occur with increasing frequency and are projected to continue along this trajectory. Crisis resilience management (CRM) plays a vital role in preparing for, mitigating, accelerating recovery for, and reducing crisis impacts on society. According to the United Nations Office for Disaster Risk Reduction (UNDRR), resilience is defined as the ability of systems (e.g., infrastructure, businesses, emergency response, and critical facilities) to resist, absorb, accommodate, adapt to, transform, and recover from the effects of a hazard in a timely and efficient manner, thereby reducing the social, economic, physical, and well-being impacts of disasters \citep{undrr_resilience_2022}. 

Information management is essential for CRM in three ways: (1) collecting and sharing critical information, e.g., to support decision-makers and residents to take preventive or preparatory actions before the hazard occurs, (2) using information, e.g., to identify and prioritize areas affected and assessing the direct and indirect impacts, and (3) analyzing information, e.g., to allocate resources equitably in response to and recovery from disasters \citep{fan_disaster_2021}. Artificial intelligence (AI) systems support crisis information management by harnessing data to provide timely decision support \citep{duan_artificial_2019}, including data analytics and machine learning (ML) or simulation models based on intelligent agents for assessment and prediction. 

The rise of AI systems in CRM is facilitated by rapid advancements in sensing technologies, increased and inexpensive data storage capacities, and increased computing speeds that enable unprecedented data collection, storage, and processing. Various emerging data, such as high-resolution satellite imagery, cell phone movement patterns, and social media data, have been integrated into AI systems for different CRM information tasks to support decision-making at different crisis phases such as (1) preparedness, e.g., predictive risk assessment \citep{yuan_smart_2022}; (2) response, e.g., rapid damage assessment \citep{lee_community-scale_2021,yuan_unveiling_2022} and population migration and evacuation \citep{lee_specifying_2022}, or (3) recovery, e.g., situational awareness \citep{zhang_semiautomated_2020}. 

Along with the increasing prominence of AI in different fields, principles and standards have been established to guide the design, development and use of AI, commonly referred to as responsible AI principles \citep{barredo_arrieta_explainable_2020}. These principles have found use in domains such as health care, security, finance, hiring and human resources, and news \citep{dwivedi_artificial_2021}. However, despite the rapid growth in the creation and adoption of AI systems in CRM, important responsible AI issues are yet to be addressed. This is especially problematic as crises have been shown to more severely impact vulnerable populations. Table \ref{tab:1} summarizes examples of various datasets in the context of CRM and their corresponding major responsible AI concerns. In the absence of addressing responsible AI issues, use of technologies may incur consequences that deteriorate CRM processes in the short- or long-term. These responsible AI issues need to be addressed and embedded into the design and development for the CRM field to take full advantage of AI systems.

By their very nature, crises create novel situations and emergent organizational arrangements that respond to them \citep{comes_coordination-information_2020}. This emergent structure poses information management challenges for data sharing and data standards, as well as for data analysis via AI systems \citep{nespeca_towards_2020}. For example, if an AI for structural damage assessment is designed as a part of crisis response, there are likely going to be incompatibility issues when this AI is adopted for another region because buildings may be of a different quality or the structures, population distribution, or infrastructure, from which the AI learned to identify structural damage, may not be the same, potentially causing the AI to over- or underestimate damage, resulting in misallocation of funding or resources. Further, information flow, decision-making authority, or privacy and data standards may be fundamentally different. The areas of poly-crises (i.e., overlapping and intersecting crises, such as COVID-19 and heatwaves) require special attention since they may create compound effects, feedback loops, or long-term effects that do not occur in isolated settings conventionally considered during the design of an AI system. 

AI systems in CRM are created to facilitate better situational awareness and decision-making by providing more information; however, in the context of crises, more information does not always translate to better decisions. For instance, in the response to Typhoon Haiyan, the humanitarian community had massively increased their capacity to collect and analyze data \citep{comes_bringing_2015}. However, multiple questions arose on the use and usefulness of the new data sources, along with confusion around which decisions or initiatives were to be supported with this data. The spiral of data collection and processing with unclear purpose or audience culminated in information overload and neglect of important information \citep{comes_coordination-information_2020}. To this date, the multiple calls for actionable, relevant, and operational information have only been partially met. One of the key complications is that in crises, there is an overwhelming number of actors and it is often unclear who will make which decision based on what information. Due to the increased data volume, data preparedness and data governance gain rising popularity in CRM. From designing to deploying AI models, organizations, researchers and developers need an in-depth understanding of data, data limitations, and possible biases. Communication among crisis managers, researchers, and data scientists regarding responsible AI issues needs to be strengthened so that there is a comprehensive understanding of different concerns and considerations of data and AI systems. 

These examples highlight the significance of addressing responsible AI issues in CRM. To address the lack of a roadmap and research agenda toward responsible AI in CRM, the main research questions addressed in this paper are: (1) What are the core responsible AI challenges in the current state of the art and current practices related to CRM? and (2) What priority research areas for responsible AI in CRM need to be addressed over the next decade? 

As one of the first attempts to address responsible AI issues in CRM, we examine different core aspects such as (1) equity and fairness; (2) biases in data, analyses, and decisions; (3) explainability and transparency; (4) accountability and credibility; (5) inter-organizational coordination and public involvement; and (6) information privacy and security. These issues and considerations correspond to different stages of the crisis information management lifecycle. In the following, we will develop responsible AI principles that are framed as propositions to describe the key challenges of each field and specify research areas that need to be further addressed in the near future.

\begin{landscape}
\begin{longtable}{>{\hspace{0pt}}m{0.17\linewidth}>{\hspace{0pt}}m{0.23\linewidth}>{\hspace{0pt}}m{0.25\linewidth}>{\hspace{0pt}}m{0.23\linewidth}}
\label{tab:1}\\
\caption{Examples of various datasets used in CRM and their major responsible AI concerns.}\\ 
\toprule
\textbf{ Datasets } & \textbf{ AI methods } & \textbf{ AI task examples } & \textbf{ Major Responsible AI concerns } \endfirsthead 
\midrule
Social media data \newline (Twitter, Facebook, etc.) & Natural language processing; Network analysis; Random forest; Support vector machines; Factorization machines. & Early warning; Damage monitoring; Infectious disease contact evaluation; Recovery evaluation; Situational awareness; Social impact assessment; Fake news detection; Sentiment analysis. & Data bias and representation; Data privacy; Information overload; Credibility and accountability; Explainability and transparency \\
\\
Satellite imagery; \newline Unmanned aerial vehicle imagery; CCTV camera data & Computer vision techniques; Deep learning techniques; Convolution neural network. & Damage assessment; Susceptibility and risk assessment; Damage monitoring; Recovery evaluation. & Data privacy; Explainability and transparency; Data security; Inter-organizational coordination \\
\\
Digital trace data; \newline Traffic data; \newline Point-of-interest data & Geospatial data analytics; Spatial and temporal statistical techniques; Network analysis; Spatiotemporal graphic neural network. & Susceptibility and risk assessment; Damage monitoring; Mobility impact assessment; Social impact assessment; Infectious disease cases prediction. & Equity/fairness consideration; Data and algorithm biases; Data sharing; Explainability and transparency; \\
\\
Mobile phone activity data; \newline Population mobility data & Graph neural network; Reinforcement learning; Long short-term memory (LSTM); Geospatial data analytics. & Damage monitoring; Crowd monitoring; Traffic prediction; Location mapping; Mobility impact assessment; Mobility evaluation; Recovery evaluation & Equity/fairness consideration; Data and algorithm biases; Data sharing; Explainability and transparency \\
\\
Crowdsourced data & Data mining techniques; Random forest; Support vector machines; Statistical analysis. & Damage monitoring; Risk prediction (e.g., flood, fire); Recovery evaluation. & Data bias and representation; Credibility and accountability; Data privacy; Data security; \\
\\
Sensor data & Spatial and temporal statistical techniques; Bayesian modeling; Regression analysis; Support vector machines. & Flood prediction; Earthquake magnitude prediction. & Credibility and accountability; Inter-organizational coordination; \\
\\
\bottomrule
\end{longtable}
\end{landscape}

\section{CRM AI Systems must promote equity and fairness considering diverse stakeholder values}
A critical aspect of creating AI systems for CRM is the consideration of equity and fairness. Equity and fairness are closely linked: equity in the context of CRM is defined as reducing the disproportionate impacts of crises on vulnerable populations \citep{berke_recovery_1993}. Fairness in the context of data science and ML is defined as AI systems whose results are independent of a given variable, especially sensitive attributes such as race and income \citep{corbett-davies_measure_2018}. Hence, establishing suitable criteria for fairness in the data analytics and ML models is necessary for achieving equity in the CRM process. The lack of consideration of various fairness criteria (such as group, individual, and causal fairness) in AI models can be greatly problematic and contribute to a lack of equity in CRM. For example, if technological solutions developed for enhanced flood monitoring and situational awareness do not consider critical fairness issues, they will widen the disparity in the impacts of disasters on vulnerable sub-populations. 

During the COVID-19 pandemic, the well-known phenomenon that existing inequalities are amplified by crises was demonstrated again. Around the world, decisions about the pandemic were informed by epidemiological models that predicted the direct impact of the pandemic in terms of the number of people infected or the strain on the public health system. Research has shown that the most vulnerable people were most exposed to the spread of the disease \citep{chang_mobility_2021}, gender gaps were widened \citep{czymara_cause_2021}, and the poorest households experienced the highest losses of household income \citep{perugini_social_2021}. In addition, school closures impeded learning and disproportionately affected disadvantaged children \citep{armitage_considering_2020}. This clearly shows that data-driven pandemic models created for informing policies did not fully account for equity considerations. At the same time that inequalities are amplified by the crisis, inequality also impacts crisis response; poverty and inequality in a community reduce the capacity for social resilience \citep{sherrieb_measuring_2010}. Because amplification of inequality is associated with a lack of trust in the authorities during the response, equity and fairness are prerequisites for an effective response.

AI systems for decision support can distort fairness and equity, as AI systems often represent and implement normative choices. Yet, in their design, it is virtually impossible to include the preferences and values of all possible and future stakeholders. Examples here are manifold, ranging from the increasing push for automated decision-making (e.g., for forecast-based finance or anticipatory action \citep{coughlan_de_perez_forecast-based_2015}) to resilience assessments for prioritization of disaster-affected regions or sectors \citep{hong_measuring_2021}, to optimization models for rapid disaster relief logistics \citep{baharmand_bi-objective_2019}. To this end, it is critical that the inclusivity and representation of all stakeholder groups is ensured so that AI systems can be designed to balance competing interests. While decision-making in crises and disasters should be guided by principles such as impartiality, neutrality, and humanity and considerations of equity and fairness \citep{van_de_walle_nature_2015}, there is still a lack of a formalization of these principles, which prevents them from being translated into AI models. Instead, current models are designed to maximize efficiency and effectiveness for crisis managers and the organizations for which they work \citep{finn_fundamentally_2016}. Thus, the challenge is for AI developers and researchers to translate moral concepts into formal methods and develop tools to support decision-makers facing critical trade-offs between (humanitarian) principles and norms, fairness, effectiveness, and efficiency

Although most models in AI systems for decision support neglect equity and fairness considerations, a notable exception is the inclusion of fairness via welfare economics principles by accounting for deprivation cost \citep{holguin-veras_appropriate_2013}. Deprivation cost is the economic value of the human suffering caused by the lack of goods or services in crises. However, these utilitarian approaches are clearly not equipped to represent moral concepts such as impartiality and humanity in AI systems related to CRM. Further, the use of deprivation costs assumes that people can adequately determine their willingness to pay prior to a crisis — even though it is well-known that crises and disasters drastically change values and preferences — and thereby also prices. Examples range from the increasing attention to public health in the COVID-19 pandemic to the turn to massively increased budgets for defense in response to the war in Ukraine. Hence, the determination of criteria and requirements for equity/fairness of AI systems would depend on the values and preferences of decision-makers and the public which could evolve in CRM. Also, CRM AI systems should inform equitable decisions, considering evolving conditions, after crises using proper fairness criteria. 

To address the considerations of equity and fairness, it is essential to understand the theoretical basis of equity and fairness and to identify conceptual analyses and components of fairness necessary for responsible AI systems in CRM. In addition, determining the roles of the principles of neutrality and impartiality in the goal of responsible AI systems is necessary. For example, during a disaster event such as a flood, decision-makers cannot simply distribute resources and supplies equally to all residents in terms of neutrality and impartiality and view it as fairness. Some people need less help because they have sufficient resources to recover from flood impacts, but others require more resources than an equal portion to prevent subsequent hazards such as those caused by bacteria and mold. AI system developers and users need to identify specific components and principles of fairness and equity to address the concerns of inequality in CRM. The understanding of equity and fairness might also vary amongst groups and resist being codified. An equity-aware approach is more likely to be trusted and used by residents and community members if AI developers successfully improve community engagement. Moreover, the benefits of fairness-aware AI systems for CRM are multifold. In addition to more equitable decisions and actions that fairness-aware CRM AI systems could enable, if technological solutions are designed to account for equity considerations, they could enhance the much-needed equity awareness and compassion in public officials, emergency managers, and responders who consume the output information.

We appeal to the AI community to (1) focus on the impact of AI systems on the most vulnerable and marginalized groups; (2) define and formalize equity/fairness criteria for CRM AI systems; (3) formalize and integrate trade-offs between norms and criteria based on the values of diverse stakeholders. The most vulnerable need to take center stage because of moral and equity considerations, but may also harbor distrust of crisis management authorities, as these institutions may be perceived to not act in their interest. Defining equity/fairness criteria will require mapping stakeholder values in relation to the decision problem for which the AI system is being developed and specify accordingly proper equity/fairness criteria. Also, issues of data justice and fairness often come to the fore when we productively move away from ideas of a universal “person” to be specific about whom we are talking. The challenges here are ones that face many institutions and have no easy solutions but turning a blind eye to the problems is not a responsible approach. AI developers can collaborate with social scientists focused on participatory action research or with ethicists who can elucidate moral concepts.

\section{CRM AI systems should facilitate bias mitigation in data, analyses, and decision-making}
Another consideration, closely related to equity and fairness, is the issue of bias in data, analyses, and decision-making. Researchers found that the complex decision-making situations in crises induce decision biases \citep{zhou_emergency_2018}. Decision makers in crises are confronted with tremendous time pressure, stress, and uncertainty: creating conditions in which decision makers tend to neglect biases in datasets and AI algorithms while amplifying biases in decision-making \citep{paulus_interplay_2022}. Accordingly, biases pertain to decisions when under risk \citep{montibeller_cognitive_2015} and during crisis response \citep{comes_cognitive_2016}. Furthermore, research on natural and conflict disasters has shown that decision makers tend to prioritize short-term responsive action, reflecting the urgency and time pressure at hand \citep{comes_coordination-information_2020}, but neglecting longer-term strategic consequences and irreversibilities. Financially, the massive bailouts and support packages after the COVID-19 crises have created path dependencies for years to come. The decision of Germany and the Netherlands to re-open coal plants in response to the energy crisis caused by the Russian invasion in Ukraine highlights that the potential longer-term consequences for the climate crises are inappropriately discounted. Because the decisions taken during crises are likely to have long-term implications, there is a clear need to counter short-sighted biases in datasets, narrow algorithms, and biased decisions.

In addition to impacts on decisions, issues such as imbalanced datasets and biased algorithms in models are particularly critical in CRM \citep{akter_algorithmic_2021,fan_spatial_2020}. For example, recent studies have shown issues related to the underrepresentation of certain populations (e.g., low-income and racial minorities) in location-based and crowdsourced datasets cause group fairness issues. Also, social media users are sometimes aware that the fact that their work is being read can have real effects on the unfolding of a disaster \citep{crawford_limits_2015}. Performances on social media can lead to coordinated misinformation campaigns \citep{starbird_disinformation_2019}. Moreover, imbalanced and underrepresented data and biased algorithms exacerbate the issues of inequalities caused by power differentials. Power is not distributed evenly throughout a population. In the US context and elsewhere, race, ethnicity, language, nationality, financial resources, location, religion, gender, sexuality, ability, and other dynamics are intimately involved with the assignment of power. But these are all macro-dynamics: within particular organizations, families, societies, and localities, there are wildly different organizations of power. AI systems in the context of CRM can create inequities when they rely on algorithms or datasets that obscure or reify power relations among groups of people. People or groups may be entirely unrepresented in datasets, leading to biased disaster resilience programs. Often the management of a crisis occurs from a distance which may hinder understanding or awareness of the nuances of local relations \citep{chouliaraki_improper_2011}. The resources available to a senior disaster manager at FEMA will be different from those available to a local or regional manager. For example, the source of risk or suffering might not be under the control of a local population, but the paradigm of resilience requires people who are suffering to live with risks \citep{finn_documenting_2018,tierney_resilience_2015,walker_genealogies_2011}. When power differentials between those who are impacted by a crisis, or between crisis managers/researchers and the people are affected are ignored or unacknowledged, it can further negatively impact fairness, creating double disasters as those who already marginalized are disproportionately affected by the disaster and then ignored in the response \citep{madianou_digital_2015,madianou_technocolonialism_2019}.

Technologies developed for CRM can have profound effects on those who receive aid. Recent research has shown that biased datasets and algorithms create path dependencies, from which decision-makers are unable to adjust their initial decisions, even though they know that the initial data was flawed \citep{paulus_interplay_2022}. Yet, there remains a widespread belief in crisis management that decisions made based on biased data or algorithms will be corrected and adjusted as more information becomes available \citep{corbacioglu_organisational_2006}. Therefore, it is incumbent on responsible AI developers to acknowledge that their technologies may create lasting bias. It is therefore paramount that responsible AI attend to power differences using participatory or other approaches to ensure democratic processes are incorporated and to identify biases introduced during data collections and model development, such as historical biases, measurement biases, and aggregation biases \citep{suresh_framework_2021}. Developers of AI models need to know as much as possible about how and why data was collected to understand gaps and biases in datasets that might impact their usage in AI models. In addition, implementing debiasing techniques during data processing and model development is a requirement to accommodate bias issues. Future research should (1) develop a comprehensive bias identification framework and metrics that can be applied to current and emerging data and algorithms; (2) track how biases propagate from data through AI to sequential and interdependent decisions; and (3) propose debiasing pipelines that attend to power differences to augment responsible AI in CRM.

\section{CRM AI systems should be explainable and transparent to gain broader trust}
Advanced AI algorithms can be useful in the context of CRM. For example, deep learning models (e.g., applications of convolution neural networks) can inform disaster response by using satellite imagery to detect areas that have been affected by a crisis \citep{gupta_rescuenet_2021}. However, many machine learning techniques, especially those relying on neural networks, create stochastic outputs that cannot easily be explained to decision makers and are often considered to be black boxes. Amidst accusations of AI systems as unfair or improper actors, many have argued that transparency is crucial to ensure that stakeholders trust machines. In the high-stake decisions typical for CRM, the inability to explain the outputs produced by models regarding disaster risks and impacts can influence decision makers’ and users’ trust in AI systems. 

Generally, the purpose of explainable AI is to make the behavior of an algorithm more intelligible by providing explanations \citep{gunning_xaiexplainable_2019}. For example, the response to COVID-19 heavily relied on voluntary compliance; transparency and thereby trust in the population were vital \citep{singh_impacts_2021}. Research has further shown that trust in the authorities and understanding of the rationale behind the decisions are vital to achieve compliance \citep{guo_warning_2022}. For instance, lack of trust in predictions has led to a failure to evacuate early, in turn causing many casualties both in the L’Acquila earthquake \citep{alexander_aquila_2010} as well as in the 2021 floods in Northwestern Europe \citep{cornwall_europes_2021}. However, in some cases, users of AI systems might put too much trust in the model outputs, neglecting to interpret the outputs in conjunction with their knowledge and risk tolerance \citep{french_believe_2005}. In other cases, the users might not trust the outputs and hence exclude the insights and foresights from their decision-making process. One approach to these problems is to increase the transparency into an AI system and ensure that AI systems that inform CRM decisions are interpretable. However, while many papers postulate a link between explainability and trust in CRM, the mechanisms behind are not yet fully understood.

Interpretable models that can reveal feature interaction and feature importance provide information for decision makers who understand how to implement AI algorithms. However, models with less interpretability (sometimes referred to as a “black box”) due to their complexity often provide better accuracy compared to more interpretable models \citep{lundberg_unified_2017}. Hence, during the early stages of AI system development, all users and stakeholders should understand trade-offs between explainability and accuracy and have a clear definition of the desired level of accuracy and interpretability. 

Explainability and transparency can create opportunities for new knowledge to improve CRM processes, but researchers have also argued that there is little reason to be hopeful about the capacity of transparency to make for predetermined sets of goals such as fairness. \citet{ananny_seeing_2018} described the limits of transparency in AI systems such as: transparency is not connected to systems of punishment or reward and so though transparency might reveal transgressions, there may be no changes as a result of the revelations; transparency can cause harm to vulnerable groups if it reveals secrets; transparency does not always produce something that is usable and may actually create more confusion; transparency doesn’t necessarily build trust in systems because trust is a social concept that is experienced differently by different people; and transparency may not help people who lack technical knowledge, or as Ananny and Crawford put it, “seeing is not understanding” \citep{ananny_seeing_2018}. 

Although transparency and explainability do not address the existential question of whether some types of AI systems in CRM should exist at all, transparency and explainability are necessary for knowledge about how CRM processes work. Thus, transparency and explainability are keys to public oversight and public input, but require other institutional apparatus, such as measures of transparency and explainability of AI systems, to make them helpful to achieve goals of responsible AI. In addition, if an AI system is opened for investigation, there have to be real consequences for the system, including that development and use of the system ceases, to ensure trust in not only the AI, but the process of public accountability. Also, while the expertise to understand AI technologies lies with a few experts who understand both the domains of AI and crises, the investigation of AI systems has to include people who are supposed beneficiaries, as well as public employees and administrators. That is, diverse publics need to be included in oversight. Explainable models are more likely to facilitate relationship building and knowledge sharing among organizations and stakeholders. Hence, to move forward, it is essential for research to (1) explicitly define transparency and explainability for AI systems in CRM, including how to measure transparency and explainability of AI systems, what needs to be transparent in AI systems, and to what extent the explainability of AI systems is acceptable; (2) understand the link between explainability, transparency, and trust; and (3) improve knowledge sharing and AI system understanding among diverse organizations to better cope with crises (something that we address further in the next section).

\section{CRM AI systems should yield credible insights for accountable decision making}
Predictive analytics is one capability provided by AI systems for crisis preparedness and response. An example of predictive AI in disaster response is when researchers use past data about rainfall and weather models to predict future flood risk — important for forecasting flood risk management decisions such as buyout programs (e.g., in the USA, this is a federal program where homeowners are compensated for property in flood prone areas) and infrastructure improvements. When decisions are informed by AI systems, there are important challenges related to accountability and credibility that need to be addressed. Accountability refers to the idea that certain people or organizations accept responsibility for the results of AI systems \citep{rasche_toward_2009}. The problem of accountability in AI is challenging because it is difficult to determine who should be responsible for the impact of an AI, and how this responsibility should be implemented or enforced. A related concept, credibility, like trust, is a property of relations where an organization, people, or technology is held in high esteem \citep{metzger_making_2007}. 

Researchers, developers, and decision makers need a clear definition of responsibilities and accountability when an AI model is created and used for CRM. However, decision makers may expect ideal predictions and thus overestimate the capabilities of models, resulting in undesirable outcomes. For example, AI systems have been adopted to predict what areas and roads might be flooded. Models make predictions about which roads will get inundated in the next few hours. However, there is a risk of false-negative and false-positive predictions. Models predicting that a road will not be flooded but then does become flooded could result in death or injury. A prediction of flooding on one road can lead to unnecessary bottlenecks elsewhere. In Indonesia, a false-negative prediction resulted in more than 1,200 people killed due to a tsunami. Even though the earthquake was detected and felt, people did not evacuate because an early warning system for tsunami detection that had credibility failed to detect and warn of three tsunami waves \citep{singhvi_what_2018}. Although a predictive model of road inundations could save hundreds of lives based on correct predictions, false-negative or false-positive predictions could result in dire outcomes. 

Ignoring the results from AI models also has dire consequences. The extent and magnitude of floods in Northwestern Europe in July 2021 was correctly predicted by the European meteorological services. But the warnings were discarded at the regional and local level in Germany as not credible, because the amounts of rain that had been predicted in a very short period were simply unimaginable to the decision makers \citep{cornwall_europes_2021}. This neglect led to delays in the order of evacuations, which resulted in dozens of casualties in the affected communities. 

In some cases, the burden of accountability can scare agencies and developers from adopting or creating AI systems. Accordingly, the issue of accountability requires a clear understanding among all entities involved and will likely change depending on the local laws. Importantly, research is needed to (1) co-create reference frameworks and standards for specific AI systems with AI developers, designers, and stakeholders to report chains of development of specific models to improve accountability; (2) educate developers, decision makers, and other model stakeholders about model limitations, underlying assumptions, possible areas for application, and model uncertainty with public documentation and innovative visualizations. For example, instead of a model predicting whether a road gets inundated or not, a model should/can provide the likelihood of road flooding. Decision makers then need training to interpret the results along with their own knowledge and risk threshold and communicate with residents. The decision makers and users should also be familiar with the way AI systems work and their inherent capabilities and limitations to avoid over- and underestimating the capabilities of models. Besides the agreed reference frameworks and standards, innovative research around visualizing uncertainty can help researchers communicate with different stakeholder groups.

\section{Inter-organizational coordination and public involvement are critical for creating responsible CRM AI systems}
Crisis Resilience Management includes diverse organizations and stakeholders. We have known for decades that information sharing amongst organizations is the backbone of effective coordination and crisis response \citep{quarantelli_disaster_1988}. For example, the Humanitarian Data Exchange, a data platform operated by the United Nations Office for the Coordination of Humanitarian Affairs (UN OCHA), provides a space for humanitarian organizations to upload and store their data for decision support. Researchers can utilize the datasets to develop AI systems. However, the sharing of crisis data among groups of collaborators brings up issues of data ownership, data stewardship, and complex issues about meta-data and interpretation \citep{finn_fundamentally_2016}. Data sharing is a significant barrier to achieving integrated AI-based solutions developed upon different datasets owned by different organizations. Yet, AI systems for CRM require the cooperation of multiple organizations in sharing data. Organizations involved with CRM can create and use their own AI systems in isolation or collaboratively depending on limitations on information sharing or organizational responsibilities and legal authorities. Using AI systems in isolation could negatively affect collective sense-making, information processing, and coordinated decision making, all of which are core elements for effective CRM \citep{quarantelli_disaster_1988}.

In addition to inter-organizational data sharing challenges, data implemented in CRM AI systems comes in a plethora of forms, and its collection ranges from using traditional methods, such as census surveys, to modern tools, such as satellite imagery, sensors, social media, GPS, and credit card transactions. Data can introduce and multiply errors to models, analyses, and interpretations, especially while merging various datasets together. In this way, decision makers are confronted with a paradoxical situation of a deluge of uncertain, noisy, conflicting information, while some key datasets may be missing \citep{altay_challenges_2014,comes_bringing_2015}. Although fusing data from various sources might introduce errors, data having finer and higher resolution and better data quality provide opportunities to improve the performances of AI models. Accordingly, data needs meta-data or stories about it to tell developers and users about its provenance and any quirks or intricacies that they should note when interpreting it. Researchers should attend to information quality requirements for crisis response organizations to improve the management of information (e.g., \citet{bharosa_challenges_2010}).

One approach to address issues of inter-organization collaboration is to create federated AI systems with human-centered AI (HCAI) frameworks based on close coordination among CRM organizations and including involvement of the stakeholders and publics influenced by an AI system \citep{aragon_human-centered_2022,shneiderman_human-centered_2022}. A federated AI system is not fully integrated; instead, it means different AI systems interface with one another and share results. An important challenge here is that given the increasing importance of bottom-up initiatives in crisis response, CRM AI systems need to be designed to be adaptive to the emerging roles and information-sharing structures that are typical for today’s crises \citep{nespeca_towards_2020}.

The task of identifying and engaging all stakeholders is perhaps one of the most challenging aspects of HCAI for CRM, since the ecosystem of CRM is already fragmented in most countries with limited coordination and cooperation among stakeholders. The HCAI process, however, could be a mechanism through which diverse stakeholders could better share data, cooperate and coordinate disaster resilience management plans and actions. In addition, HCAI can urge more communications between the AI developers and the organizations adopting AI systems. This could enable developers to make systems that better speak to users, and for users to learn more about the limitations of the systems and how to interpret results. 

Creating an integrated federated AI framework requires close coordination among various organizations. Integrating AI models created in isolation into a federated AI framework is challenging in the absence of institutional connectedness. In governance settings where institutional connectedness is absent, individual AI-based models for crisis resilience are created by organizations in isolation, and the combined effects of models on plans and actions would be less than the sum of the parts. However, federating organizations might also create privacy challenges (that we discuss in more detail below). Federated AI frameworks, however, could be created so that models learning from private datasets can communicate outputs with each other. 

Another important aspect of inter-organizational collaboration is peer review and validation of methods underlying AI systems in CRM. Including processes of peer review increases trust among organizations. However, some technologies may not be peer reviewed or validated because of commercial and intellectual property interests. Prioritizing the protection of private interests could be problematic in CRM since the models may not be tested until a crisis strikes and the model results are not trusted by collaborating organizations.

A collaborative ecosystem of researchers, developers, publics, crisis managers, humanitarian agencies, and other stakeholders should be formed to peer review and validate whether CRM AI systems are aligned with the required processes and standards. A federated AI system with human-centered AI frameworks could require incentive mechanisms and funding sources that provide incentives for coordination and facilitate open innovation. To enhance inter-organizational coordination and public involvement, future research should (1) develop a framework to facilitate federated AI systems and human-centered AI processes; (2) develop a standard to ensure data quality and its corresponding details (e.g., limitations) when sharing amongst various organizations; and (3) propose incentive mechanisms to encourage inter-organizational coordination and public involvement with peer-review and validation to augment responsible AI in CRM.

\section{CRM AI systems should attend to information privacy and security}
Major drivers of AI systems in CRM are the advancements in sensing technologies at a higher resolution that can include considerable personal data and other data that requires attention to information security and privacy. The massive collection of personal data about human behavior, ranging from mobility data collected via mobile phones to video footage from CCTV cameras and or imagery collected via UAVs, inevitably has led to many questions about the ownership and the principles and protocols to deal with this data, and the requirements that guide its use. Different legal regimes govern the use of this data by AIs, notably the EU’s General Data Protection Regulation (GDPR) puts restrictions on data access and reuse. Additionally, ethical concerns about the use of data collected during or before disaster events for emergency response purposes include whether individuals can have control of their information. In many cases, because of the scale and urgency of the emergency, the data collection and use happen without informed consent. An important argument for the massive collection of personal data, especially in crises, is the urgency of the situation. Parallels can be drawn to emergency medicine, whereby privacy can be violated and data about health status is shared if it serves the survival of the patient. Following this argument, some say that data collection via such methods as aerial imagery can only be justified only if it directly and immediately serves the population about which the data is collected. 

Informed consent has to include the option to opt out. Often, beneficiaries affected by a crisis or disaster are dependent on assistance, and they hardly have the free choice to opt out. In times of disaster, people may be at the worst moments of their lives — do AI system developers have the moral authority to make use of their social media in these moments \citep{finn_documenting_2018}? This is amplified by the increasing use of biometric technologies to identify beneficiaries around the globe \citep{jacobsen_experimentation_2015}; while one can abandon a cash voucher or ID card, this is impossible with fingerprints or iris scans. Further, there is the right to be forgotten, a feature of the GDPR in the EU, which implies that individuals can ask for the removal of their data. This may, however, be difficult if the data is published widely via dashboards or social media \citep{starbird_could_2016}. Moreover, individuals may not even be aware that their data are used for specific analyses or information products \citep{panger_reassessing_2016,shilton_excavating_2021}. Therefore, appropriate mechanisms need to be implemented that allow individuals to understand where and for which purposes their data is being used, and to guarantee that the data can be withdrawn at any moment. 

Besides privacy concerns, data sharing is subjected to important security considerations. Malicious actors might introduce fabricated and misleading data. While there are many advocates for open and public data sharing in crisis, there are important pitfalls to consider, especially in conflicts or human-made crises. Data in conflicts is especially sensitive since malicious actors can strategically exploit seemingly innocent data to target the most vital infrastructures of society, such as hospitals, schools, bakeries, or humanitarian convoys. Datasets such as individual-level mobility data have sensitive information about populations and should not be shared. Therefore, even the collection, processing, and sharing of information that is considered “public” in a natural disaster have to be considered carefully in conflicts and should follow the idea of minimizing information flows, along with clear data standards, security levels, and data sharing protocols. Creations of accessible data archives that respect security and privacy interests are paramount, which ensures a chain of data stewardship that protects the integrity and quality of the data.

Different contexts require AI system developers to have different moral sensitivities, the ability to recognize moral issues and to understand moral consequences of decisions. For example, the moral sensitivity and concerns that are needed for an AI system for advertising are different from those in the context of crisis response. Consequently, unique ethical considerations are required in the context of CRM that can inform issues around information privacy and security. However, there is little agreement on the best ethical approach to CRM. Historically, the ethics that guide disaster response is contested. Some actors might be guided by rights-based frameworks such as the capabilities approach, adopted by many development organizations(e.g., \citet{nussbaum_womens_2000,sen_commodities_1999}) while others focus on the crisis standards of care \citep{leider_ethical_2017} enacted by MSF, an international and independent medical humanitarian organization. In our view, the focus of using responsible AI in CRM should be especially focused on the affected populations and their needs. Even with agreement on an ethics framework to guide CRM, implementing ethical codes in AI systems is challenging if not impossible \citep{lewis_global_2020,sun_mapping_2019}, and no guarantee that an AI system will be ethical. Thus, AI system developers need to talk to different stakeholders to agree on an ethics approach that is grounded in shared, articulated principles and they might consider implementing a process for addressing ethical dilemmas and penalties for violating the ethics frameworks. Enacting a public, ethical approach to AI for CRM is important for trust and credibility as discussed above.

To address privacy and security concerns, in addition to creating an ethics framework to guide work, it is necessary to build upon coordination and cooperation across researchers, public, and private stakeholders, as discussed above. Regulations and ethics codes are needed. In terms of data protection, the EU has developed the GDPR, which is widely regarded as the gold standard in data protection. It is legally required and critical to protect personal data or at least data which may identify individuals or specific ethnicities or groups in the context of CRM. Future research can focus on three important directions: (1) creating datasets that respect for users’ rights and privacy regulations; (2) innovative approaches to address privacy concerns; and (3) addressing data provenance and stewardship in a way that respects security and privacy. Researchers can work with groups such as TrustedCI to develop strategies for ensuring information security and with archives such as Inter-university Consortium for Political and Social Research (ICPSR) at the University of Michigan or other reputable public archives that have thoughtful data policies and stewards.

\section{Discussion}
\subsection{Reflection on theoretical gaps, framework and research proposition development}
The number of studies and tools related to AI in CRM has grown considerably over the past few years; however, the majority of the existing literature is missing elements of responsible AI. Several studies \citep{duan_artificial_2019, dwivedi_artificial_2021} have discussed the role of AI in supporting decision-making, yet the discussion of AI related to CRM is very limited. Because decision-making in crises and disasters can significantly affect human life and create irreversible impacts for our societies and environment, AI used to support CRM must be designed to maintain principles such as impartiality, neutrality, and humanity and considerations of equity and fairness. Building upon the discussion of AI in supporting decision-making, we further specify AI used in the context of crises and disasters based on the information management lifecycle, including data collection, data analysis, data sharing, and decision-making.

In this paper, we presented a roadmap with six propositions to address responsible AI issues in CRM:\\
Proposition 1- \textit{CRM AI Systems must promote equity and fairness considering diverse stakeholder values}: It is essential to define, theorize, and formalize criteria for incorporating equity and fairness in AI systems to support CRM. The most vulnerable populations should be considered when implementing AI to support decision-making in the context of crises and disasters.

Proposition 2- \textit{CRM AI systems should facilitate bias mitigation in data, analyses, and decision-making}: Issues such as imbalanced datasets and biased algorithms are critical to CRM since they can result in undesirable outcomes. Since data biases cannot be avoided in the urgency of crises, a comprehensive bias identification and mitigation framework is necessary for CRM AI systems to avoid detrimental results.

Proposition 3- \textit{CRM AI systems should be explainable and transparent to gain broader trust}: The explainability and transparency of AI systems and their results can influence trust in AI systems. A key challenge in CRM is making complex problems explainable under time pressure. To this end, it is necessary to explicitly define and empirically measure transparency and explainability for AI systems in CRM, and understand how transparency and explainability propagate.

Proposition 4- \textit{CRM AI systems should yield responsible insights for accountable decision making}: While decision-making during crises and disasters has enormous impacts, it is essential to clearly define the extent of accountability of AI developers, researchers, and decision makers when an AI model is created and used for CRM. Communications among all stakeholders regarding issues such as model limitations, assumptions and uncertainties are critical to ensure people don’t over- or under-estimate the capability of AI systems.

Proposition 5- \textit{Inter-organizational coordination and public involvement are critical for creating responsible CRM AI systems}: There are diverse organizations and stakeholders involved in CRM. Identifying and engaging all stakeholders during AI design and development, despite challenging, they are critical to promote information sharing and coordination of plans and actions.

Proposition 6- \textit{CRM AI systems should attend to information privacy and security}: Privacy and security should not be sacrificed due to the urgency and severity of crisis events; instead, rules and ethics frameworks encourage information privacy and security during data sharing and implementation. It is important to establish regulations and encourage innovative approaches to ensure privacy and security during times of crises while harnessing emerging data that could be beneficial for informing CRM decisions and actions.

\subsection{Contributions to research/theoretical implications}
\begin{figure}
	\centering
    \includegraphics[width=0.9\linewidth]{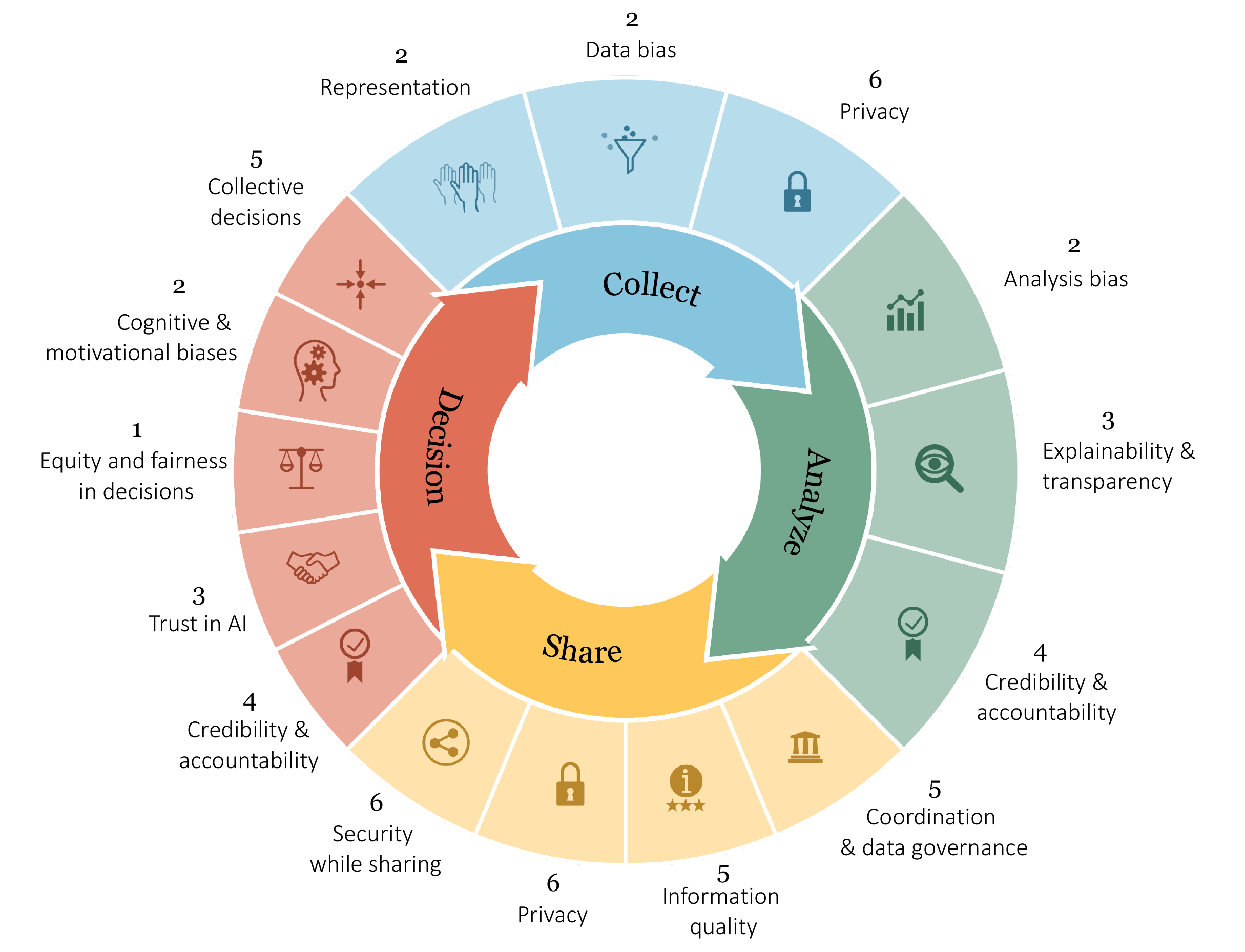}
    \caption{Responsible AI challenges in the Crisis Resilience Information Management lifecycle. The numbers indicate the proposition we address regarding the responsible AI challenges.}
	\label{fig:fig1}
\end{figure}

Despite growing research related to AI in CRM, very limited attention has been paid to responsible AI practices in crisis information management. The crisis information management cycle consists of collecting, analyzing, sharing information, and making decisions based on information \citep{van_de_walle_nature_2015}. Through identifying challenges, specifying propositions, and mapping these onto existing models of crisis information management, the contribution of this study is to address important theoretical challenges for information management research, that span across the fields of information system and AI design, cognition and informational use, as well as ethics and normative questions. The six propositions for responsible AI in CRM that are organized according to the crisis information management cycle are a roadmap to fill this gap. In Figure \ref{fig:fig1}, we show how the propositions for responsible AI in CRM map on to theoretical models of crisis information management focused on cycles of information collection, analysis, sharing, and decision-making.

Our intervention elaborates on the crisis information management processes and identifies normative sub-dimensions of how these responsible AI processes map onto these processes. For example, during the analysis phase of crisis information management, we show that responsible AI practices mean that this phase includes: preventing or mitigating analysis bias (Proposition 2), ensuring explainability and transparency of analysis for stakeholders (Proposition 3), and articulating chains of accountability for making use of AI analysis (Proposition 4). In addition, because of the cognitive, behavioral and moral dimensions of crisis information management, we stress that human-centered design is paramount across all phases of the cycles. Figure \ref{fig:fig1} depicts the research challenges we put forward and need to be addressed to establish the foundations of responsible AI research in CRM. The roadmap and the propositions specified in this study serve as the first necessary step to bring responsible AI issues in CRM to the forefront of research agenda in this growing field. Further studies, discussed below in Section \ref{sec:sec8_4}, are needed to establish frameworks and practices for incorporating responsible AI considerations in CRM and to implement the propositions we put forward in this paper. We demonstrate how designing systems and analyzing data with AI responsibly fit in the context of crisis information management.

\subsection{Implications for practice}
The propositions specified in this study are intended to promote responsible AI in CRM in practice and to improve AI systems in supporting CRM decision-making. This study has broad implications for AI in CRM practice, ranging from developers to emergency managers, humanitarian organizations, and public officials. A prerequisite to any implementation or formulation of standards and guidance is understanding the theoretical basis of responsible AI issues and collectively defining conceptual ideas related to responsible AI concerns. There is a need to measure and formalize standards around equity and fairness and to explicitly define transparency and explainability for AI systems in CRM. In addition, understanding the mechanisms behind the relationships between explainability, transparency, trust, as well as issues around accountability is vital for AI design to develop AI systems that can be trusted by decision makers and therefore be effective during crises and disasters.

We summarize the practical implications of the responsible AI propositions discussed in this study for AI developers, decision makers, and public officials:

(1) Implications for AI developers\\
The purpose of AI systems in CRM is to facilitate better informed decisions and actions. Yet, CRM decisions are known to be fraught with a range of cognitive challenges driven by urgency and high stakes. To account for these aspects in human decision-making, following the propositions specified in this study can facilitate more human-centered and user-centered design of AI systems in CRM. For example, greater interpretability, transparency, and explainability can facilitate greater trust in AI systems, fostering, in turn, the integration of AI into the decision-making processes. In addition, including bias identification and mitigation framework and the consideration of equity and fairness in AI systems can avoid detrimental AI outcomes and subsequent decisions. Lack of consideration of fairness and biases in CRM AI systems and subsequent decisions is greatly problematic since crises disproportionately impact vulnerable populations. The propositions specified in this study provide a roadmap that needs attention to promote responsible AI system design and development in CRM.

(2) Implications for decision makers\\
Making decisions under tremendous time pressure and uncertainty is known to induce cognitive biases, which may result in over- or under-estimating the capacity of AI systems and discarding biases in datasets and AI algorithms. To mitigate this effect, it is necessary to involve decision-makers in AI design. This step will also draw attention to the chains of development, in turn enabling accountability and credibility. At the same time, co-design will help developers and decision-makers understand and improve the AI systems they work with as a prerequisite for joint standards. The issues of fairness and equity in AI systems are particularly critical in CRM decision-making. Responsible AI system co-design has the potential to reveal the blind spots in CRM decisions, plans, and policies that lead to inequity in crisis impacts among vulnerable populations to support decision-making in CRM. The approach we proposed requires decision makers to collaborate with researchers, developers, and the public with different skillsets and perspectives to facilitate better coordination among different stakeholders. 

(3) Implications for public officials\\
Since the issues and considerations discussed in this study correspond to different stages of the crisis information management lifecycle, it is necessary to (1) create ethics frameworks to guide the development of AI systems and (2) develop standards, guidance, and regulations with incentive mechanisms and consequences for the AI system developers, AI users, and decision makers. With the ethics frameworks and regulations, AI system development might be reorganized such that AI systems address the responsible AI propositions discussed in this study. Although the frameworks and regulations may seem burdensome, they can guide the development and use of AI systems in CRM and ensure AI systems perform better with responsible AI considerations. The propositions specified in this study could result in the development of frameworks and regulations that include responsible AI considerations with coordination among different stakeholders, researchers, and developers involved in creating AI systems for CRM.

\subsection{Limitations and future research direction}
\label{sec:sec8_4}
We put forward the six propositions based on literature reviews, authors’ experiences, and discussion with experts and practitioners in the fields. The issues discussed here are not comprehensive but include the most significant elements that we (a group of interdisciplinary researchers with diverse expertise) have collectively encountered during several years of research in this area and in close interactions and discussions with various public, private and humanitarian organizations. The issues that we identify are interwoven as highlighted in our discussions throughout the paper. For instance, fairness issues are associated with issues of bias and transparency, and issues of transparency influence inter-organizational coordination and collective decision making. Each of the ideas (e.g., fairness or transparency) discussed in this paper could be further explored on its own and open various lines of inquiries. Table \ref{tab:2} summarizes the recommended future research directions discussed in this paper toward responsible AI in CRM. Future case studies are also needed to systematically and rigorously validate, test, and refine the proposed roadmap with the six propositions with further empirical investigations.

\begin{longtable}{>{\hspace{0pt}}m{0.288\linewidth}>{\hspace{0pt}}m{0.648\linewidth}}
\label{tab:2}\\
\caption{Summary of future research directions toward responsible AI in CRM.}\\ 
\toprule
\textbf{Aspects of responsible AI issues} & \textbf{Future research directions} \endfirsthead 
\midrule
CRM AI Systems must promote equity and
  fairness considering diverse stakeholder values & (1) understand the theoretical basis of equity and fairness in the context of CRM. \par{}(2) define and formalize equity/fairness criteria for CRM AI systems based  \par{}(3) integrate trade-offs between different criteria based on the values of diverse stakeholders. \\
\\
CRM AI systems should facilitate bias mitigation in data, analyses, and decision making\textbf{} & (1) develop a comprehensive bias identification framework and metrics that can be applied to current and emerging data and algorithms. \par{}(2) track how biases propagate from data through AI to sequential and interdependent decisions. \par{}(3) propose debiasing pipelines that attend to power differences to augment responsible AI in CRM.\textbf{} \\
\\
CRM AI systems should be explainable and transparent to gain broader trust\textbf{} & (1) explicitly define transparency and explainability for AI systems in CRM. \par{}(2) understand the link between explainability, transparency, and trust. \par{}(3) improve knowledge sharing and AI system understanding among diverse organizations to better cope with crises.\textbf{} \\
\\
CRM AI systems should yield credible insights for accountable decision making\textbf{} & (1) co-create standards for specific AI systems with AI developers, designers, and stakeholders to report chains of development of specific models to improve accountability. \par{}(2) develop a framework to educate developers, decision makers, and other model stakeholders about model limitations, underlying assumptions, possible areas for application, and model uncertainty with public documentation and innovative visualizations.\textbf{} \\
\\
Inter-organizational coordination and public involvement are critical for creating responsible CRM AI systems\textbf{} & (1) develop a framework to facilitate federated AI systems and human-centered AI processes. \par{}(2) develop a standard to ensure data quality and its corresponding details (e.g., limitations) when sharing amongst various organizations. \par{}(3) propose incentive mechanisms to encourage inter-organizational coordination and public involvement with peer-review and validation to augment responsible AI in CRM.\textbf{} \\
\\
CRM AI systems should attend to information privacy and security\textbf{} & (1) create datasets that include helpful information and respect for users’ rights and privacy regulations. \par{}(2) develop innovative approaches to address privacy concerns. \par{}(3) address data provenance and stewardship in a way that respects security and privacy.\textbf{} \\
\bottomrule
\end{longtable}
\section{Conclusions}
This paper is intended to bring the urgent issue of responsible AI to the forefront and to formulate a clear set of propositions to direct the attention of the interdisciplinary community of researchers across data and information sciences, engineering, social sciences, humanities, geography, urban science, and disaster science to the most pressing challenges. In addition, we aim to raise awareness among practitioners to trigger discussions and research to find approaches and solutions to effectively address the issues around responsible AI in CRM. In particular, we highlight the issues related to responsible AI in the context of the complex and dynamic decision-making environments typical for crises including equity, fairness, biases, explainability and transparency, accountability, privacy and security, inter-organizational coordination, and public engagement to demonstrate the significance of the topic and the urgency of addressing these issues. The directions for future studies toward responsible AI in CRM are outlined in this paper. In addition, systematic and rigorous case studies are required to further investigate and validate the roadmap proposed.

\bibliography{ref}  






\end{document}